\documentclass[aps,prc,showpacs,amssymb,amsmath,amsfonts,floatfix,nofootinbib]{revtex4-2}

\usepackage[inner=2.25cm,outer=2.25cm,top=1.0cm,bottom=2cm,includeheadfoot]{geometry}

\makeatletter
\def\p@subsection{}
\def\p@subsubsection{}
\makeatother

\usepackage{amsmath}
\usepackage[usenames]{color}
\usepackage{epsfig}
\usepackage{epstopdf}
\usepackage{amssymb,amsmath}
\usepackage{amsmath}
\usepackage{subfig}
\usepackage{epstopdf}
\usepackage{sidecap}
\usepackage{cases}
\usepackage{enumitem}
\usepackage{bm}
\usepackage{overpic}
\usepackage{upgreek}
\usepackage{morefloats}
\usepackage{xcolor}
\usepackage{listings}

\definecolor{green}{rgb}{0,1.0,0}
\definecolor{gray}{rgb}{0.4,0.4,0.4}
\definecolor{black}{rgb}{0,0,0}

\newcommand{\be}{\begin{eqnarray}}
\newcommand{\ee}{\end{eqnarray}}
\newcommand{\bc}{\begin{center}}
\newcommand{\ec}{\end{center}}
\newcommand{\beq}{\begin{eqnarray}}
\newcommand{\eea}{\end{eqnarray}}

\newcommand{\Ocal}{\mathcal{O}}

\newcommand{\UCONN}{University of Connecticut, Storrs, CT 06269, USA}

\newcommand{\JLab}{Thomas Jefferson National Accelerator Facility, Newport News, VA 23606, USA;}

\begin{document}

\allowdisplaybreaks

\title{General integral-identities involving Legendre polynomials and their derivatives}
\author{Yannick~Wunderlich}\email[Corresponding author: ]{yannick.wunderlich@uconn.edu}\affiliation{\UCONN} 
\author{Kyungseon~Joo}\affiliation{\UCONN}\author{Victor~I.~Mokeev}\affiliation{\JLab}

\date{\today}

\begin{abstract}
Integrals involving derivatives of Legendre polynomials frequently arise in applications ranging from multipole expansions for processes involving electromagnetic probes (cf.~\cite{BallPhotoproduction,KDT}) to spectral methods in numerical physics. Despite their practical relevance, closed-form expressions for such integrals - particularly involving arbitrary derivative orders - are not readily accessible in standard references or symbolic tools. In this note, we derive and present general analytic expressions for integrals of the form
\begin{equation}
  \int_{-1}^{+1} dx P^{(q)}_{n} (x) P^{(k)}_{m} (x)  , \nonumber 
\end{equation}
where $P_{n} (x)$ and $P_{m} (x)$ are Legendre polynomials and $q$, $k$ denote their order of differentiation. Using repeated integration by parts, parity arguments, and closed-form boundary evaluations, we obtain explicit binomial and Gamma-function representations valid for all non-negative integers $n$, $m$, $q$, $k$. These results unify and extend known orthogonality relations and provide ready-to-use tools for analytic and computational contexts.
\end{abstract}

\maketitle

\tableofcontents

\section{Introduction} \label{sec:Intro}

Integrals involving Legendre polynomials and their derivatives arise in a broad range of physical and mathematical contexts - from multipole expansions in classical electrodynamics, quantum angular momentum theory and partial-wave analysis (see e.g. formalisms in~\cite{BallPhotoproduction,KDT}), to modern applications in spectral methods and orthogonal polynomial techniques for numerical solutions of differential equations. Despite their prevalence in such domains, closed-form expressions for integrals involving derivatives of Legendre polynomials - especially at arbitrary differentiation orders - are surprisingly absent from standard references. \\

Classical resources such as Abramowitz \& Stegun~\cite{AbramowitzStegun}, the Digital Library of Mathematical Functions~\cite{dlmf}, and comprehensive integral tables like Gradshteyn \& Ryzhik~\cite{gradryzhik} or Prudnikov, Brychkov \& Marichev~\cite{prudnikov} typically provide orthogonality relations and product integrals involving the polynomials themselves, but offer only limited results for derivative overlaps. The work~\cite{QureshiEtAl} has considered special cases, particularly involving first- and second-order derivatives, but these tend to focus on either symmetric cases (e.g., $\int P_n'(x)P_n'(x)\,dx$) or specific recursion-based identities. More recent treatments, such as~\cite{kholshevnikov}, explore integrals of iterated Legendre functions or related generating functions, but still do not provide a general framework for integrals of the form
\begin{equation}
  \int_{-1}^{+1} dx P^{(q)}_{n} (x) P^{(k)}_{m} (x)  , \label{eq:GeneralIntegralIntro}
\end{equation}
where $q$, $k$ are arbitrary non-negative integers denoting derivative orders. \\

This note aims to provide a unified and accessible derivation of such integrals (for all non-negative integers $n, m, q, k$) by employing a combination of integration by parts, parity arguments, boundary evaluations, and factorial/Gamma-function identities. As a key ingredient, we present a general formula for the boundary values $P_n^{(k)}(x)\big|_{x = +1}$, namely
\begin{equation}
P_n^{(k)}(x) \big|_{x = +1} = \frac{(n + k)!}{2^k\,k!\,(n - k)!},
\label{eq:BoundaryValuekthDerivativeIntro}
\end{equation}
which we derive using only an elementary recurrence relation and a known uniqueness-argument from discrete mathematics (cf. Appendix~\ref{sec:BoundaryValueDeriv} of this note). While the resulting expression may be known in some form, we are not aware of any equally direct and pedagogically motivated derivation in the standard literature. Our goal is not to claim originality of the final results obtained in this work, but to present them in a form that is compact, verifiable, and useful for both analytical and numerical applications.

\section{Overlap of ordinary Legendre polynomial with 1st order derivative} \label{sec:1stOrderDeriv}

We begin by proving the following result for the integral of a Legendre polynomial times the derivative of another Legendre polynomial:
\begin{equation}
 \int_{-1}^{+1} dx P_{n} (x) P'_{m} (x) = \begin{cases}  0, & n+m \text{ even,} \\ 0, & n+m \text{ odd and } m < n, \\ 2, & n+m \text{ odd and } n < m .\end{cases}     \label{eq:LegPolIntRelation}
\end{equation}
\textit{Proof:} \\
Via integration by parts, we can show relatively quickly that:
\begin{equation}
 \int_{-1}^{+1} dx P_{n} (x) P'_{m} (x) = \left[ P_{n} (x) P_{m} (x) \right] \big|_{-1}^{+1} - \int_{-1}^{+1} dx P'_{n} (x) P_{m} (x)   . \label{eq:IntByPartsProofI}
\end{equation}
Now, at the boundaries $x = \pm 1$, it is well-known that Legendre polynomials satisfy (due to definite parity of the polynomials, as well as normalization at $x = +1$):
\begin{equation}
 P_{n} (+1) = 1 \text{, and } P_{n} (-1) = (-1)^{n} . \label{eq:LegPolynomialsBoundaryValues}
\end{equation}
Thus, our integral becomes
\begin{equation}
 \int_{-1}^{+1} dx P_{n} (x) P'_{m} (x) = \left[ 1 - (-1)^{m+n} \right]  - \int_{-1}^{+1} dx P'_{n} (x) P_{m} (x)   . \label{eq:IntByPartsProofII}
\end{equation}
For $m+n$ even, we have $\left[ 1 - (-1)^{m+n} \right] = 0$ and therefore
\begin{equation}
 \int_{-1}^{+1} dx P_{n} (x) P'_{m} (x) =  - \int_{-1}^{+1} dx P'_{n} (x) P_{m} (x)   . \label{eq:IntByPartsProofIII}
\end{equation}
Now to show that for this case, our considered integral is indeed zero, we have to consider further parity-arguments. Start with
\begin{equation}
 P_{n} (-x) = (-1)^{n} P_{n} (x) . \label{eq:BasicLegPolParity} 
\end{equation}
For the derivative, we obtain the following due to the chain rule:
\begin{equation}
 P'_{n} (-x) = \frac{d}{d (-x)} P_{n} (-x) = \frac{dx}{d(-x)} \frac{d}{dx} (-1)^{n} P_{n} (x) = (-1)^{n+1} P'_{n}(x) . \label{eq:DerivLegPolParity} 
\end{equation}
Thus, for our integrand $f(x) := P_{n} (x) P'_{m} (x)$, the parity relation becomes:
\begin{equation}
  f(-x) = P_{n} (-x) P'_{m} (-x) = (-1)^{n} P_{n}(x) (-1)^{m+1} P'_{m} (x) = (-1)^{m+n+1} f(x)  , \label{eq:IntegrandParity}
\end{equation}
and we see directly that for even values of $m+n$, our integrand is odd and thus the integral vanishes. \\
Now, let us return to our general relation~\eqref{eq:IntByPartsProofI} and assume that on the left-hand-side there, we have $m+n$ odd and $n<m$ satisfied. Thus, we get 
\begin{equation}
 \int_{-1}^{+1} dx P_{n} (x) P'_{m} (x) = \left[ 1 - (-1)^{m+n} \right] - \underbrace{\int_{-1}^{+1} dx P'_{n} (x) P_{m} (x)}_{m+n \text{ odd and } n<m \text{ here}}   . \label{eq:IntByPartsProofNew}
\end{equation}
Since the Legendre polynomials form a complete orthogonal system of functions, we know that for the derivative $P'_{n} (x)$ under the integral in the right-hand-side, the following re-expansion formula holds:
\begin{equation}
 P'_{n} (x) = \sum_{j = 0}^{n-1} a^{(n-1)}_{j} P_{j} (x)    , \label{eq:DerivGeneralReExp}
\end{equation}
with some as of yet unspecified coefficients $a^{(n-1)}_{j}$. We know that a general expansion of the form~\eqref{eq:DerivGeneralReExp} has to exist and terminate at the order $\Ocal (x^{n-1})$, since the derivative $P'_{n} (x)$ is a polynomial of the same order.
However, the re-expansion~\eqref{eq:DerivGeneralReExp} into Legendre-polynomials without derivatives involves only ordinary Legendre polynomials $P_{j} (x)$ of order $j < n$. Thus, we also have $j < m$ for all $j$, and the integral appearing on the right-hand-side of equation~\eqref{eq:IntByPartsProofNew} vanishes, due to basic and well-known orthogonality properties of Legendre polynomials. We furthermore get
\begin{equation}
 \underbrace{\int_{-1}^{+1} dx P_{n} (x) P'_{m} (x)}_{_{m+n \text{ odd and } n<m \text{ here}}} = \left[ 1 - (-1)^{m+n} \right] = 1 - (-1) = 2 , \label{eq:IntByPartsProofNewII}
\end{equation}
i.e. the integral is in this case fully fixed by the boundary term '$\left[ 1 - (-1)^{m+n} \right]$'. This completes our derivation of the general result~\eqref{eq:LegPolIntRelation}. \hfill \textbf{QED.} \\

The result~\eqref{eq:LegPolIntRelation} can also be written in a more concise form using the Heaviside $\theta$-function:
\begin{equation}
 \int_{-1}^{+1} dx P_{n} (x) P'_{m} (x) = \theta (m-n) \left[ 1 - (-1)^{m+n} \right]  .   \label{eq:LegPolIntRelationConcise}
\end{equation}
We notice that keeping the bracket '$\left[ 1 - (-1)^{m+n} \right]$' ensures a 'parity-filter'. \\

We have to declare here that within this note, we use the so-called 'right-continuous' Heaviside-function, i.e.
\begin{equation}
 \theta (x) \equiv \Theta_{>} (x) = \begin{cases} 1, & x > 0 , \\ 0, & x \leq 0 . \end{cases}    , \label{eq:RightContStepFunct}
\end{equation}
such that boundary-cases are excluded.

\section{Overlap of ordinary Legendre polynomial with 2nd order derivative} \label{sec:2ndOrderDeriv}

In many cases, one would also need the integral involving the second derivative of a Legendre-polynomial:
\begin{equation}
 \int_{-1}^{+1} dx P_{n} (x) P''_{m} (x)  . \label{eq:Int2ndDerivative}
\end{equation}
Here, one can again do integration by parts, but now two successive times:
\begin{align}
 \int_{-1}^{+1} dx P_{n} (x) P''_{m} (x) &= \left[ P_{n} (x) P'_{m} (x) \right] \big|_{-1}^{+1} - \int_{-1}^{+1} dx P'_{n} (x) P'_{m} (x)  \nonumber \\
 &=  \left[ P_{n} (x) P'_{m} (x) \right] \big|_{-1}^{+1} - \left[ P'_{n} (x) P_{m} (x) \right] \big|_{-1}^{+1} + \int_{-1}^{+1} dx P''_{n} (x) P_{m} (x)  . \label{eq:IntByParts2ndDerivProofI}
\end{align}
For the evaluation of the boundary term, one has to know that Legendre's differential equation yields (this is a special case of the more generally formula~\eqref{eq:BoundaryValuekthDerivative}, which is re-derived in Appendices~\ref{sec:BoundaryValueDeriv} and~\ref{sec:BoundaryValueDerivGenFunc})
\begin{equation}
 P'_{n} (x = +1) = \frac{n(n+1)}{2}   , \label{eq:DerivLegPolEndpoint}
\end{equation}
and therefore one also gets
\begin{equation}
 P'_{n} (x = -1) = (-1)^{n+1} P'_{n} (x = +1) = (-1)^{n+1} \frac{n(n+1)}{2}   . \label{eq:DerivLegPolEndpointII}
\end{equation}
With all this, we get for the first summand in the boundary-term
\begin{equation}
  \left[ P_{n} (x) P'_{m} (x) \right] \big|_{-1}^{+1} = 1 \cdot \frac{m(m+1)}{2} - (-1)^{n} \cdot (-1)^{m+1} \frac{m(m+1)}{2}  = \frac{m(m+1)}{2} \left[ 1 - (-1)^{n+m+1} \right]   , \label{eq:2ndDerivBoundaryTerm}
\end{equation}
and for the second one similarly
\begin{equation}
  \left[ P'_{n} (x) P_{m} (x) \right] \big|_{-1}^{+1} =  \frac{n(n+1)}{2} \cdot 1 - (-1)^{n+1} \frac{n(n+1)}{2} \cdot (-1)^{m}  = \frac{n(n+1)}{2} \left[ 1 - (-1)^{n+m+1} \right]   , \label{eq:2ndDerivBoundaryTermII}
\end{equation}
We thus effectively consider the integral:
\begin{equation}
 \int_{-1}^{+1} dx P_{n} (x) P''_{m} (x) = \left( \frac{m(m+1)}{2} - \frac{n(n+1)}{2} \right) \left[ 1 - (-1)^{n+m+1} \right] + \int_{-1}^{+1} dx P''_{n} (x) P_{m} (x)   . \label{eq:IntByParts2ndDerivProofII}
\end{equation}
We now repeat the parity-argument from above, but with a new integrand $\tilde{f} (x) := P_{n} (x) P''_{m} (x) $, which satisfies the parity-relation:
\begin{equation}
 \tilde{f} (-x) = P_{n} (-x) P''_{m} (-x) = (-1)^{n} P_{n} (x) \cdot (-1)^{m+2} P''_{m} (x) = (-1)^{n+m} \tilde{f} (x) . \label{eq:ParityRel2ndDerivIntegral} 
\end{equation}
This directly tells us that our considered integral vanishes for \textit{odd values of} $m+n$. Furthermore, we automatically get the result
\begin{equation}
 \underbrace{\int_{-1}^{+1} dx P_{n} (x) P''_{m} (x)}_{_{m+n \text{ even and } n \geq (m-1) \text{ here}}} = 0 , \label{eq:IntByPartsProofNewII}
\end{equation}
due to 're-expansion arguments' for the Legendre-polynomial $P''_{m} (x)$ under the integral (similar to arguments given below equation~\eqref{eq:IntByPartsProofNew} in section~\ref{sec:1stOrderDeriv}). Therefore, in our further considerations, we consider the case where $n < (m-1)$ and again can do this kind of 'trick' where the boundary-term effectively fixes the value of the integral, since the remaining integral on the right-hand-side vanishes:
\begin{align}
 \underbrace{\int_{-1}^{+1} dx P_{n} (x) P''_{m} (x)}_{m+n \text{ even and } n < (m-1)} &= \left( \frac{m(m+1)}{2} - \frac{n(n+1)}{2} \right) \underbrace{\left[ 1 - (-1)^{n+m+1} \right]}_{= 1 - (-1) = 2} + \underbrace{\underbrace{ \int_{-1}^{+1} dx P''_{n} (x) P_{m} (x)}_{m+n \text{ even and } n < (m-1)}}_{=0} \nonumber \\
 &= \left[  m(m+1) - n(n+1)  \right]  . \label{eq:IntByParts2ndDerivProofIII}
\end{align}
We assemble everything we have learned about the integral~\eqref{eq:Int2ndDerivative} in the following form:
\begin{equation}
 \int_{-1}^{+1} dx P_{n} (x) P''_{m} (x) = \begin{cases}  0, & n+m \text{ odd,} \\ 0, & n+m \text{ even and } n \geq (m-1), \\ \left[  m(m+1) - n(n+1)  \right], & n+m \text{ even and } n < (m-1) .\end{cases}     \label{eq:2ndLegPolIntResult}
\end{equation}
This result can also written more concisely as:
\begin{equation}
 \int_{-1}^{+1} dx P_{n} (x) P''_{m} (x) = \theta ([m-1]-n) \left[ 1 - (-1)^{n+m+1} \right] \left( \frac{m(m+1)}{2} - \frac{n(n+1)}{2} \right) .    \label{eq:2ndLegPolIntResultConcise}
\end{equation}

\section{Generalization to overlap with k-th order derivative} \label{sec:kthOrderDeriv}

It is worth noting that the present method can be used in order to derive the generalized overlap-integral with a general $k$-th derivative of a Legendre polynomial:
\begin{equation}
 \int_{-1}^{+1} dx P_{n} (x) P^{(k)}_{m} (x) \text{, where } P^{(k)}_{m} (x) \equiv \frac{d^{k}}{d x^{k}} P_{m} (x) \text{, and } k \geq 1 . \label{eq:IntkthDerivative}
\end{equation}
As a first step, performing integration-by-parts $k$ consecutive times (this approach is also central to the work in~\cite{QureshiEtAl}) can be used in order to 'shuffle' all derivatives from the polynomial $P_{m}$ to $P_{n}$, i.e. we expect an intermediate result of the form:
\begin{equation}
 \int_{-1}^{+1} dx P_{n} (x) P^{(k)}_{m} (x) = \left[ \text{'boundary-terms'} \right] + (-1)^{k} \int_{-1}^{+1} dx P^{(k)}_{n} (x) P_{m} (x)  . \label{eq:IntByPartskthDerivProofI}
\end{equation}
We will specify the boundary-terms later on, since they become a bit more involved. First, let us note that the integrand $f(x) := P_{n} (x) P^{(k)}_{m} (x)$ satisfies the parity-relation
\begin{equation}
 f(-x) =  P_{n} (-x) P^{(k)}_{m} (-x) = (-1)^{n} P_{n} (x) (-1)^{m+k} P^{(k)}_{m} (x) = (-1)^{n+m+k} f(x)  . \label{eq:ParityRelGeneralIntegrand}
\end{equation}
This directly gives us the case for vanishing of our integral, namely:
\begin{equation}
 \int_{-1}^{+1} dx P_{n} (x) P^{(k)}_{m} (x) = 0 \text{, for } n+m+k \text{ odd}. \label{eq:GeneralCaseVanishingCondition}
\end{equation}
Furthermore, the 're-expansion-argument' used repeatedly above (cf. sections~\ref{sec:1stOrderDeriv} and~\ref{sec:2ndOrderDeriv}) directly gives us:
\begin{equation}
 \int_{-1}^{+1} dx P_{n} (x) P^{(k)}_{m} (x) = 0 \text{, for } n \geq m - (k-1) . \label{eq:GeneralCaseReExpansionCondition}
\end{equation}
Therefore, in the only case where the original integral does \textit{not} vanish, it is again given by the boundary terms, due to:
\begin{align}
 \underbrace{\int_{-1}^{+1} dx P_{n} (x) P^{(k)}_{m} (x)}_{_{n+m+k \text{ even and } n < m - [k-1] \text{ here}}} &= \left[ \text{'boundary-terms'} \right] + (-1)^{k} \underbrace{\underbrace{\int_{-1}^{+1} dx P^{(k)}_{n} (x) P_{m} (x)}_{_{n+m+k \text{ even and } n < m - [k-1] \text{ here}}}}_{=0} \nonumber \\
 &=  \left[ \text{'boundary-terms'} \right]  . \label{eq:IntByPartskthDerivProofII}
\end{align}
The next ingredient needed for the evaluation of our general result are the generalized 'boundary-terms', which appeared first in equation~\eqref{eq:IntByPartskthDerivProofI}. Generalizing the results found in sections~\ref{sec:1stOrderDeriv} and~\ref{sec:2ndOrderDeriv}, we can write:
\begin{align}
  \left[ \text{'boundary-terms'} \right] &= \left[ P_{n} (x) P^{(k-1)}_{m} (x) \right] \Big|_{-1}^{+1} - \left[ P'_{n} (x) P^{(k-2)}_{m} (x) \right] \Big|_{-1}^{+1} + \left[ P''_{n} (x) P^{(k-3)}_{m} (x) \right] \Big|_{-1}^{+1} \nonumber \\
  & \hspace*{12pt} + \ldots + (-1)^{k-2} \left[ P^{(k-2)}_{n} (x) P'_{m} (x) \right] \Big|_{-1}^{+1} + (-1)^{k-1} \left[ P^{(k-1)}_{n} (x) P_{m} (x) \right] \Big|_{-1}^{+1} \nonumber \\
  &=  \sum_{j=1}^{k} (-1)^{j-1} \left[ P^{(j-1)}_{n} (x) P^{(k-j)}_{m} (x) \right] \Big|_{-1}^{+1} . \label{eq:BoundaryTermGeneralized}
\end{align}
The only missing ingredient necessary for the final result for our general integral~\eqref{eq:IntkthDerivative} is the boundary-value at $x = +1$ for the general $k$-th derivative $P^{(k)}_{m} (x)$ of a Legendre polynomial. The derivation of the resulting formula is non-trivial, but it can be done either using recurrence relations, or via generating functions for Legendre polynomials. The proof involving recurrence relations is rather pretty and maybe not standard, so we describe it in some detail in Appendix~\ref{sec:BoundaryValueDeriv}. For completeness, we also outline in Appendix~\ref{sec:BoundaryValueDerivGenFunc} a generating-function based proof of the boundary-value identity. While this approach is classical and well-known, including it here serves to emphasize the robustness of the result and may be more familiar to some readers. The resulting expression for the boundary-value reads:
\begin{equation}
  P^{(k)}_{m} (x) \big|_{x = +1} = \frac{(k+m)!}{2^{k} k! (m - k)!}  . \label{eq:BoundaryValuekthDerivative}
\end{equation}
The first few cases for lowest $k$, for this general expression, read:
\begin{align}
 P_{m} (x) \big|_{x = +1} &= 1 , \label{eq:DerivBoundaryValueExampleI} \\
 P'_{m} (x) \big|_{x = +1} &= \frac{m(m+1)}{2} , \label{eq:DerivBoundaryValueExampleII} \\
 P''_{m} (x) \big|_{x = +1} &= \frac{(m-1)m(m+1)(m+2)}{8} , \label{eq:DerivBoundaryValueExampleIII} \\
 P'''_{m} (x) \big|_{x = +1} &= \ldots . \label{eq:DerivBoundaryValueExampleIV} 
\end{align}
Using this knowledge, we can evaluate our boundary-terms to:
\begin{align}
 \left[ \text{'boundary-terms'} \right]  &= \sum_{j=1}^{k} (-1)^{j-1} \left[ P^{(j-1)}_{n} (x) P^{(k-j)}_{m} (x) \right] \Big|_{-1}^{+1} \nonumber \\
 &= \sum_{j=1}^{k} (-1)^{j-1} P^{(j-1)}_{n} (1) P^{(k-j)}_{m} (1) \left[ 1 - (-1)^{n+(j-1)+m+(k-j)} \right] \nonumber \\
 &= \left[ 1 - (-1)^{n+m+k-1} \right] \sum_{j=1}^{k} (-1)^{j-1} P^{(j-1)}_{n} (1) P^{(k-j)}_{m} (1) \nonumber \\
 &=  \left[ 1 - (-1)^{n+m+k-1} \right] \sum_{j=1}^{k} (-1)^{j-1} \frac{([j-1]+n)!}{2^{(j-1)} (j-1)! (n - [j-1])!} \frac{([k-j]+m)!}{2^{(k-j)} (k-j)! (m - [k-j])!} . \label{eq:BoundaryTermsGeneralResult}
\end{align}
In the third step, we notice that the bracket, which we pulled in front of the sum, acts as a 'parity-filter' for our final result, effectively enforcing the rule~\eqref{eq:GeneralCaseVanishingCondition} (cf. the result~\eqref{eq:GeneralIntkthDerivativeFinalResult} further below). 
Since the determination of the boundary-terms has been accomplished, the integral is known as well, just as in the simpler cases discussed in earlier sections. \\
We can implement the 'inequality-constraints' for the (non-) vanishing of the integral using a Heaviside $\theta$-function (cf. our convention defined in equation~\eqref{eq:RightContStepFunct}). With this, our final result becomes:
\begin{align}
   \int_{-1}^{+1} dx P_{n} (x) P^{(k)}_{m} (x) &= \theta \left( \left\{ m - (k-1) \right\} - n \right) \left[ 1 - (-1)^{n+m+k-1} \right] \nonumber \\
  & \hspace*{12pt} \times \frac{1}{2^{(k-1)}} \sum_{j=1}^{k} (-1)^{j-1} \frac{([j-1]+n)!}{ (j-1)! (n - [j-1])!} \frac{([k-j]+m)!}{ (k-j)! (m - [k-j])!} . \label{eq:GeneralIntkthDerivativeFinalResult}
\end{align}
As can be checked, this equation reduces to the results already obtained in section~\ref{sec:1stOrderDeriv} for $k=1$, and in section~\ref{sec:2ndOrderDeriv} for $k=2$, i.e.:
\begin{align}
 \int_{-1}^{+1} dx P_{n} (x) P^{(k=1)}_{m} (x) &= \theta \left( \left\{ m - (1-1) \right\} - n \right) \left[ 1 - (-1)^{n+m+1-1} \right] \nonumber \\
  & \hspace*{12pt} \times \frac{1}{2^{(1-1)}} \sum_{j=1}^{1} (-1)^{j-1} \frac{([j-1]+n)!}{ (j-1)! (n - [j-1])!} \frac{([1-j]+m)!}{ (1-j)! (m - [1-j])!} \nonumber \\
  &=  \theta \left( m  - n \right) \left[ 1 - (-1)^{n+m} \right] , \label{eq:GeneralIntkthDerivativeLimitI} \\
  \int_{-1}^{+1} dx P_{n} (x) P^{(k=2)}_{m} (x) &= \theta \left( \left\{ m - (2-1) \right\} - n \right) \left[ 1 - (-1)^{n+m+2-1} \right] \nonumber \\
  & \hspace*{12pt} \times \frac{1}{2^{(2-1)}} \sum_{j=1}^{2} (-1)^{j-1} \frac{([j-1]+n)!}{ (j-1)! (n - [j-1])!} \frac{([2-j]+m)!}{ (2-j)! (m - [2-j])!} \nonumber \\
  &=  \theta \left( [m - 1]  - n \right) \left[ 1 - (-1)^{n+m+1} \right] \frac{1}{2} \left( \frac{n!}{n!} \frac{(m+1)!}{(m-1)!} - \frac{(n+1)!}{(n-1)!} \frac{m!}{m!} \right) \nonumber \\
  &=  \theta \left( [m - 1]  - n \right) \left[ 1 - (-1)^{n+m+1} \right] \left( \frac{m (m+1)}{2} - \frac{n (n+1)}{2} \right)  . \label{eq:GeneralIntkthDerivativeLimitII}
\end{align}

We can further change the appearance of this result using the $\Gamma$-function, via
\begin{equation}
  \Gamma(n) = (n-1)! \Leftrightarrow n! \equiv \Gamma(n+1)   . \label{eq:GammaFctDef}
\end{equation}
Then, we can write our result in a form which might be suitable for analytic continuation in $(n,m;k)$, as soon as one has figured out what to do with the $\theta$-function, i.e.:
\begin{align}
   \int_{-1}^{+1} dx P_{n} (x) P^{(k)}_{m} (x) &= \theta \left( \left\{ m - (k-1) \right\} - n \right) \left[ 1 - e^{i \pi (n+m+k-1)} \right] \nonumber \\
  & \hspace*{12pt} \times \frac{1}{2^{(k-1)}}\sum_{j=1}^{k} e^{i \pi (j-1)} \frac{ \Gamma (j+n)}{ \Gamma(j) \Gamma (n-j+2)} \frac{ \Gamma([k-j]+m+1)}{  \Gamma(k-j+1) \Gamma(m - [k-j] + 1)} . \label{eq:GeneralIntkthDerivativeFinalResultGammaFct}
\end{align}
The Heaviside function itself is of course not analytic, but might be approximated as the limit of some analytic function. Then, the derived integral might become useful for Regge-theory or other related subjects, which make heavy use of analyticity reasoning.

\section{Overlap between two 1st-order differentiated Legendre-polynomials} \label{sec:Two1stOrderDeriv}

In this section, we want to discuss in more detail the case where one has two 1st-order differentiated Legendre polynomials under the integral:
\begin{equation}
 \int_{-1}^{+1} dx P'_{n} (x) P'_{m} (x)  .    \label{eq:LegPolTwo1stIntRelation}
\end{equation}
This integral can be quickly obtained, building on previously-derived results. Integration by parts yields:
\begin{equation}
  \int_{-1}^{+1} dx P'_{n} (x) P'_{m} (x) = \left[ P_{n} (x) P'_{m} (x)  \right] \Big|_{-1}^{+1} - \int_{-1}^{+1} dx P_{n} (x) P''_{m} (x)  . \label{eq:LegPolTwo1stIntStepI}
\end{equation}
The boundary term is fixed via the boundary-values of derivatives of Legendre polynomials we discussed before (cf. equation~\eqref{eq:BoundaryValuekthDerivative} and Appendices~\ref{sec:BoundaryValueDeriv} and~\ref{sec:BoundaryValueDerivGenFunc}). The integral which remains on the right-hand-side has been evaluated in section~\ref{sec:2ndOrderDeriv}. Thus, we get:
\begin{align}
 \int_{-1}^{+1} dx P'_{n} (x) P'_{m} (x) &= \frac{m(m+1)}{2} \left[ 1 - (-1)^{n+m+1} \right] \nonumber \\
 & \hspace*{12pt} - \theta ([m-1]-n) \left[ 1 - (-1)^{n+m+1} \right] \left( \frac{m(m+1)}{2} - \frac{n(n+1)}{2} \right) \nonumber \\
 &= \left[ 1 - (-1)^{n+m+1} \right] \left\{ \frac{m(m+1)}{2} \left[ 1 - \theta ([m-1]-n) \right]  + \frac{n(n+1)}{2} \theta ([m-1]-n) \right\}  .  \label{eq:LegPolTwo1stIntResult}
\end{align}
This result puts us in the position to derive the most general case for two differentiated Legendre polynomials, which will be done in the next section.

\section{Overlap between two differentiated Legendre-polynomials, with general differentiation-orders $q$ and $k$} \label{sec:Generalqthkth1stOrderDeriv}

We want to extend previous results to the case where one has two general differentiated Legendre polynomials under the integral, with integration-orders $q$ and $k$, respectively:
\begin{equation}
 \int_{-1}^{+1} dx P^{(q)}_{n} (x) P^{(k)}_{m} (x)  .    \label{eq:LegPolTwoQthKthIntRelation}
\end{equation}
Performing integration by parts a total of $q$ times, one gets:
\begin{align}
 \int_{-1}^{+1} dx P^{(q)}_{n} (x) P^{(k)}_{m} (x)  &=  \left[ P^{(q-1)}_{n} (x) P^{(k)}_{m} (x) \right] \Big|_{-1}^{+1} - \left[ P^{(q-2)}_{n} (x) P^{(k+1)}_{m} (x) \right] \Big|_{-1}^{+1} + \left[ P^{(q-3)}_{n} (x) P^{(k+2)}_{m} (x) \right] \Big|_{-1}^{+1} \nonumber \\
  & \hspace*{12pt} + \ldots + (-1)^{q-2} \left[ P'_{n} (x) P^{(k+q-2)}_{m} (x) \right] \Big|_{-1}^{+1} + (-1)^{q-1} \left[ P_{n} (x) P^{(k+q-1)}_{m} (x) \right] \Big|_{-1}^{+1} \nonumber \\
  & \hspace*{12pt} + (-1)^{q} \int_{-1}^{+1} dx P_{n} (x) P^{(k+q)}_{m} (x) \nonumber \\
  &=  \sum_{j=1}^{q} (-1)^{(j-1)} \left[ P^{(q-j)}_{n} (x) P^{(k+[j-1])}_{m} (x) \right] \Big|_{-1}^{+1} + (-1)^{q} \int_{-1}^{+1} dx P_{n} (x) P^{(k+q)}_{m} (x) .  \label{eq:LegPolTwoQthKthIntRelationStepI}
\end{align}
The boundary-terms can now again be evaluated under usage of our knowledge of boundary-values of (derivatives of) Legendre polynomials (cf. equation~\eqref{eq:BoundaryValuekthDerivative})
\begin{align}
  &\sum_{j=1}^{q} (-1)^{(j-1)} \left[ P^{(q-j)}_{n} (x) P^{(k+[j-1])}_{m} (x) \right] \Big|_{-1}^{+1} = \sum_{j=1}^{q} (-1)^{(j-1)} \left\{ P^{(q-j)}_{n} (1) P^{(k+[j-1])}_{m} (1) - P^{(q-j)}_{n} (-1) P^{(k+[j-1])}_{m} (-1) \right\} \nonumber \\
  &= \sum_{j=1}^{q} (-1)^{(j-1)}  \frac{([q-j]+n)!}{2^{[q-j]} [q-j]! (n - [q-j])!}  \frac{([k+(j-1)]+m)!}{2^{[k+(j-1)]} [k+(j-1)]! (m - [k+(j-1)])!} \left[ 1 - (-1)^{n + (q-j) + m + k + [j-1]} \right] \nonumber \\
  &= \frac{1 - (-1)^{n + m + k + q - 1}}{2^{k+q-1}} \sum_{j=1}^{q} (-1)^{(j-1)} \frac{(q-j+n)!}{(q-j)! (n - q+j)!}  \frac{(k+j-1+m)!}{ (k+j-1)! (m - k - j + 1)!}  . \label{eq:LegPolTwoQthKthBoundaryTerms}
\end{align}
Using this intermediate result, our overall integral becomes (using the general integral~\eqref{eq:GeneralIntkthDerivativeFinalResult} derived in section~\ref{sec:kthOrderDeriv}):
\begin{align}
 &\int_{-1}^{+1} dx P^{(q)}_{n} (x) P^{(k)}_{m} (x)  = \frac{ 1 - (-1)^{n+m+k+q-1} }{2^{k+q-1}} \Bigg\{  \sum_{j=1}^{q} (-1)^{j-1} \frac{(q-j+n)!}{(q-j)! (n - q+j)!}  \frac{(k+j-1+m)!}{ (k+j-1)! (m - k - j + 1)!} \nonumber \\
 & \hspace*{12pt} + (-1)^{q} \theta \left( \left\{ m - (k+q-1) \right\} - n \right)  \sum_{j=1}^{k+q} (-1)^{j-1} \frac{(j-1+n)!}{ (j-1)! (n - j+1)!} \frac{(k+q-j+m)!}{ (k+q-j)! (m - k - q + j)!}  \Bigg\} .  \label{eq:LegPolTwoQthKthIntRelationStepII}
\end{align}
If we evaluate this expression for $q=0$, thus dropping the first sum inside the curly brackets, we get:
\begin{align}
  &\int_{-1}^{+1} dx P^{(q=0)}_{n} (x) P^{(k)}_{m} (x)  =  \frac{ 1 - (-1)^{n+m+k+0-1} }{2^{k+0-1}} \Bigg\{ 0 + (-1)^{0} \theta \left( \left\{ m - (k+0-1) \right\} - n \right) \nonumber \\
  & \hspace*{12pt} \times  \sum_{j=1}^{k+0} (-1)^{j-1} \frac{(j-1+n)!}{ (j-1)! (n - j+1)!} \frac{(k+0-j+m)!}{ (k+0-j)! (m - k - 0 + j)!}  \Bigg\} \nonumber \\
  &= \theta \left( \left\{ m - (k-1) \right\} - n \right) \frac{[1 - (-1)^{n+m+k-1}]}{2^{(k-1)}} \sum_{j=1}^{k} (-1)^{j-1} \frac{(j-1+n)!}{ (j-1)! (n - j+1)!} \frac{(k-j+m)!}{ (k-j)! (m - k + j)!} 
\end{align}  
This is precisely our result~\eqref{eq:GeneralIntkthDerivativeFinalResult} obtained previously in section~\ref{sec:kthOrderDeriv}. \\

Furthermore, evaluating the general result~\eqref{eq:LegPolTwoQthKthIntRelationStepII} for the special case of $(q,k) = (1,1)$, we can reproduce the expression~\eqref{eq:LegPolTwo1stIntResult} derived in section~\ref{sec:Two1stOrderDeriv}, i.e.:
\begin{align}
 &\int_{-1}^{+1} dx P^{(q=1)}_{n} (x) P^{(k=1)}_{m} (x)  = \frac{ 1 - (-1)^{n+m+1+1-1} }{2^{1+1-1}} \Bigg\{  \sum_{j=1}^{1} (-1)^{j-1} \frac{(1-j+n)!}{(1-j)! (n - 1+j)!}  \frac{(1+j-1+m)!}{ (1+j-1)! (m - 1 - j + 1)!} \nonumber \\
 & \hspace*{12pt} + (-1)^{1} \theta \left( \left\{ m - (1+1-1) \right\} - n \right)  \sum_{j=1}^{1+1} (-1)^{j-1} \frac{(j-1+n)!}{ (j-1)! (n - j+1)!} \frac{(1+1-j+m)!}{ (1+1-j)! (m - 1 - 1 + j)!}  \Bigg\} \nonumber \\
 &= \frac{ 1 - (-1)^{n+m+1} }{2} \left\{ \frac{n!}{n!} \frac{(m+1)!}{(m-1)!} - \theta \left(  m - 1 - n \right) \left[ \frac{n!}{n!} \frac{(m+1)!}{(m-1)!} - \frac{(n+1)!}{(n-1)!} \frac{m!}{m!} \right] \right\} \nonumber \\
 &= \frac{ 1 - (-1)^{n+m+1} }{2} \left\{ m (m+1) - \theta \left(  m - 1 - n \right) \left[  m (m+1) - n (n+1)  \right] \right\}    .  \label{eq:LegPolTwoQthKthIntRelationExampleOneOne}
\end{align}

In the following, we give some exact values for the integral~\eqref{eq:LegPolTwoQthKthIntRelation} for the example-case of $(q,k) = (10,3)$, for which the expression~\eqref{eq:LegPolTwoQthKthIntRelationStepII} becomes:
\begin{align}
 &\int_{-1}^{+1} dx P^{(10)}_{n} (x) P^{(3)}_{m} (x)  = \frac{ 1 - (-1)^{n+m} }{2^{12}} \Bigg\{  \sum_{j=1}^{10} (-1)^{j-1} \frac{(10-j+n)!}{(10-j)! (n - 10 +j)!}  \frac{(j+2+m)!}{ (j+2)! (m - j - 2)!} \nonumber \\
 & \hspace*{12pt} + \theta \left( \left\{ m - 12 \right\} - n \right)  \sum_{j=1}^{13} (-1)^{j-1} \frac{(j-1+n)!}{ (j-1)! (n - j+1)!} \frac{(13 -j+m)!}{ (13 - j)! (m + j - 13)!}  \Bigg\} .  \label{eq:LegPolTwoQthKthIntRelationExampleCases}
\end{align}
Furthermore, as can be checked:
\begin{align}
 \int_{-1}^{+1} dx P^{(10)}_{10} (x) P^{(3)}_{3} (x)  &= 19 \hspace*{0.75pt} 641 \hspace*{0.75pt} 872 \hspace*{0.75pt} 250  , \label{eq:ExtremeExampleCaseI} \\
 \int_{-1}^{+1} dx P^{(10)}_{10} (x) P^{(3)}_{5} (x)  &= 137 \hspace*{0.75pt} 493 \hspace*{0.75pt} 105 \hspace*{0.75pt} 750  , \label{eq:ExtremeExampleCaseII} \\
 \int_{-1}^{+1} dx P^{(10)}_{11} (x) P^{(3)}_{4} (x)  &=  962 \hspace*{0.75pt} 451 \hspace*{0.75pt} 740 \hspace*{0.75pt} 250 . \label{eq:ExtremeExampleCaseIII}
\end{align}

\section{Summary \& Conclusions} \label{sec:Summary}

In this note, compact and fully explicit expressions for integrals involving derivatives of Legendre polynomials, of the general form $\int_{-1}^{1} P_n^{(q)}(x) P_m^{(k)}(x)  dx$ (with at least one of the two differentiation-orders $q$, $k$ non-vanishing), have been derived
under clearly stated parity- and degree-constraints. To our knowledge, such expressions appear to be rarely compiled in closed form in the readily accessible physics literature, despite their frequent appearance in applied contexts ranging from angular-momentum expansions to higher-order multipole projections.

A central technical ingredient was a general boundary formula for $P_n^{(k)}(x)\big|_{x=+1}$, which we derived from a basic derivative identity using a recurrence relation and a uniqueness argument from discrete mathematics. While the resulting factorial expression for $P_n^{(k)}(x) \big|_{x=+1}$ does exist in scattered sources, this elementary derivation appears to be underdocumented and may be of independent interest.

The recurrence-based derivation and the compact integrals presented here are intended to serve as a practical resource for both analytical calculations and computational implementations involving Legendre polynomials and their derivatives. \\

Possible continuations and extensions of this work include:
\begin{itemize}
\item[-] Adapting the methods presented here to the case of associated Legendre functions $P_\ell^m(x)$ and their boundary values, which are relevant for spherical harmonics and multipolar expansions on the sphere;
\item[-] Generalizing the integration techniques to cover integrals involving weight functions (e.g., $(1 - x^2)^\alpha$), or other classical orthogonal polynomial families (e.g., Chebyshev, Jacobi, Gegenbauer);
\item[-] Exploring numerical applications, such as the precomputation of inner-product matrices involving derivative bases in spectral or finite-element methods.
\end{itemize}

\begin{acknowledgments}
The work of Y.W. was supported by the United States Department of Energy during the writing of this manuscript.
\end{acknowledgments}

\appendix

\section{Derivation of boundary-values $P^{(k)}_{m} (x) \big|_{x = +1}$ for general differentiation order $k$, using recurrence relations} \label{sec:BoundaryValueDeriv}

In this appendix, we provide a self-contained derivation of the general boundary-formula
\begin{equation}
P_n^{(k)}(x) \big|_{x = +1} = \frac{(n + k)!}{2^k\,k!\,(n - k)!},
\label{eq:BoundaryValuekthDerivativeAppendix}
\end{equation}
which gives the $k$-th derivative of the Legendre polynomial $P_n(x)$ at $x = 1$, for all integers $n \geq 0$ and $0 \leq k \leq n$ (cf. equation~\eqref{eq:BoundaryValuekthDerivative} of the main text).

\noindent

Our foundation is the following well-known derivative recurrence-relation for Legendre polynomials:
\begin{equation}
P_n'(x) = n\,P_{n-1}(x) + x\,P_{n-1}'(x).
\label{eq:PnDerivRec}
\end{equation}
Differentiating both sides $k-1$ additional times and using the generalized Leibnitz-rule, we obtain:
\begin{align}
 P^{(k)}_{n}(x) &= n P^{(k-1)}_{n-1}(x) + \frac{d^{k-1}}{dx^{k-1}} \left[x P'_{n-1}(x) \right] \nonumber \\
 &= n P^{(k-1)}_{n-1}(x) + \sum_{j = 0}^{k-1} \binom{k-1}{j} \left( \frac{d^{k-1-j}}{dx^{k-1-j}} x \right) \frac{d^{j}}{dx^{j}} P'_{n-1} (x) . \label{eq:RecursionDerivationI}
\end{align}
From the sum in the second term on the right-hand-side, only the two terms for which the derivative acting on the single power of $x$ is non-vanishing, contribute. Thus, we get:
\begin{align}
 P^{(k)}_{n}(x) &= n P^{(k-1)}_{n-1}(x) + x \frac{d^{k-1}}{dx^{k-1}} P'_{n-1} (x) +
  \binom{k-1}{[k-1] - 1} \frac{d^{k-2}}{dx^{k-2}} P'_{n-1} (x)  \nonumber \\
 &= n P^{(k-1)}_{n-1}(x) + x P^{(k)}_{n-1} (x) + (k - 1) P^{(k-1)}_{n-1} (x) = (n+k-1) P^{(k-1)}_{n-1} (x) + x P^{(k)}_{n-1} (x) . \label{eq:RecursionDerivationII}
\end{align}
Evaluating this result at $x = 1$ yields the recurrence relation
\begin{equation}
P_n^{(k)}(1) = P_{n-1}^{(k)}(1) + (n + k - 1)\,P_{n-1}^{(k-1)}(1),
\label{eq:BoundaryDerivRecursion}
\end{equation}
which holds for all $n \geq 1$ and $1 \leq k \leq n$. We will use this recurrence relation to prove quite generally that formula~\eqref{eq:BoundaryValuekthDerivativeAppendix} is correct, in two consecutive steps: \\


(i) \underline{\textit{Verification of the factorial solution:}} \\

We now verify that the closed-form expression~\eqref{eq:BoundaryValuekthDerivativeAppendix} satisfies the recurrence relation~\eqref{eq:BoundaryDerivRecursion}. We define
\begin{equation}
F_n^{(k)} := \frac{(n + k)!}{2^k\,k!\,(n - k)!} , \label{eq:FnkDefAppendix}
\end{equation}
and compute:
\begin{align}
F_{n-1}^{(k)} + (n + k - 1)\,F_{n-1}^{(k - 1)}
&= \frac{(n - 1 + k)!}{2^k\,k!\,(n - 1 - k)!}
+ (n + k - 1) \hspace*{0.5pt} \frac{(n + k - 2)!}{2^{k - 1}\,(k - 1)!\,(n - k)!} \nonumber \\
&= \frac{(n + k - 1)!}{2^k\,k!\,(n - k - 1)!}
+ 2 k \hspace*{0.5pt} \frac{(n + k - 1)!}{2^k\,k!\,(n - k)!}  \nonumber \\
&= \frac{(n + k)!}{2^k\,k!\,(n - k)!} \left[ \frac{n-k}{n+k} + \frac{2 k}{n+k}  \right]  \nonumber \\
&= \frac{(n + k)!}{2^k\,k!\,(n - k)!} = F_n^{(k)}. \label{eq:FactorialSolProofAppendix}
\end{align}
This confirms that $F_n^{(k)}$ indeed solves the recurrence relation~\eqref{eq:BoundaryDerivRecursion}. \\

(ii) \underline{\textit{Initial conditions and uniqueness:}} \\

The recurrence relation~\eqref{eq:BoundaryDerivRecursion} is a linear inhomogeneous two-dimensional difference-equation on the triangular domain $\{(n, k) \in \mathbb{N}_0^2 : 0 \leq k \leq n\}$. To uniquely determine a solution, it suffices to specify boundary values. These are:

\begin{itemize}
\item[-] $P_n^{(0)}(1) = P_n(1) = 1$ for all $n$, a well-known result on Legendre polynomials we assume as known here;
\item[-] $P_n^{(k)}(1) = 0$ for all $k > n$, since $P_n(x)$ is a polynomial of degree $n$.
\end{itemize}

Given these boundary conditions, and the recurrence relation~\eqref{eq:BoundaryDerivRecursion}, the values of $P_n^{(k)}(1)$ for all $n, k$ with $0 \leq k \leq n$ are determined uniquely by discrete iteration. This is a standard result in the theory of difference equations (see, e.g., references~\cite{graham, stanley}).

We have thus shown that the factorial expression~\eqref{eq:BoundaryValuekthDerivativeAppendix} satisfies the recurrence relation~\eqref{eq:BoundaryDerivRecursion}, under the initial conditions specified above. It follows that this expression gives the unique solution for the boundary values $P_n^{(k)}(1)$ over the entire triangular region $0 \leq k \leq n$.

\section{Supplementary derivation of boundary-values $P^{(k)}_{m} (x) \big|_{x = +1}$ for general differentiation order $k$, using generating functions} \label{sec:BoundaryValueDerivGenFunc}

We provide a self-contained derivation of the general boundary-formula
\begin{equation}
P_n^{(k)}(x) \big|_{x = +1} = \frac{(n + k)!}{2^k\,k!\,(n - k)!},
\label{eq:BoundaryValuekthDerivativeAppII}
\end{equation}
which gives the $k$-th derivative of the Legendre polynomial $P_n(x)$ at $x = 1$, for all integers $n \geq 0$ and $0 \leq k \leq n$. We intend to use the following well-known generating function for Legendre polynomials:
\begin{equation}
  \frac{1}{\sqrt{1 - 2 x t + t^{2}}} = \sum_{n=0}^{\infty} P_{n} (x) t^{n}  .  \label{eq:GenFunctLegPolys}
\end{equation}
This generating function implies the following relation for the $k$-th derivative of Legendre-polynomials, applied by differentiating the above-given definition~\eqref{eq:GenFunctLegPolys} $k$-times with respect to $x$:
\begin{equation}
  \sum_{n=0}^{\infty} P^{(k)}_{n} (x) t^{n} = \frac{\partial^{k}}{\partial x^{k}} \frac{1}{\sqrt{1 - 2 x t + t^{2}}} . \label{eq:GenFunctDerivRelationI}
\end{equation}
We have to know the partial derivative on the right-hand-side. First, we compute some example cases:
\begin{align}
 \frac{\partial}{\partial x} \frac{1}{\sqrt{1 - 2 x t + t^{2}}}  &= \left(- \frac{1}{2} \right) \frac{(-2t)}{(1 - 2 x t + t^{2})^{3/2}} = \frac{t}{(1 - 2 x t + t^{2})^{3/2}} , \label{eq:PartDerivExamplesI} \\
 \frac{\partial^{2}}{\partial x^{2}} \frac{1}{\sqrt{1 - 2 x t + t^{2}}}  &= \left(- \frac{3}{2} \right) \frac{(-2t) t}{(1 - 2 x t + t^{2})^{5/2}} = \frac{3 t^{2}}{(1 - 2 x t + t^{2})^{5/2}} , \label{eq:PartDerivExamplesII} \\
 \frac{\partial^{3}}{\partial x^{3}} \frac{1}{\sqrt{1 - 2 x t + t^{2}}}  &= \left(- \frac{5}{2} \right) \frac{(-2t) 3 t^{2}}{(1 - 2 x t + t^{2})^{7/2}} = \frac{15 t^{3}}{(1 - 2 x t + t^{2})^{7/2}} , \label{eq:PartDerivExamplesIII} \\
 \frac{\partial^{4}}{\partial x^{4}} \frac{1}{\sqrt{1 - 2 x t + t^{2}}}  &= \ldots , \label{eq:PartDerivExamplesIV}
\end{align}
We extract/anticipate the general pattern:
\begin{equation}
  \frac{\partial^{k}}{\partial x^{k}} \frac{1}{\sqrt{1 - 2 x t + t^{2}}} = \frac{(2k-1)!! \hspace*{1pt} t^{k}}{\sqrt{1 - 2 x t + t^{2}}^{2k+1}}  , \label{eq:PartDerivGeneralPattern}
\end{equation}
where the 'double factorial' is defined as:
\begin{equation}
  n!! = n \times (n-2) \times (n-4) \times (\ldots)  . \label{eq:DefDoubleFactorial}
\end{equation}
Next, we evaluate the relation~\eqref{eq:GenFunctDerivRelationI} at $x=1$ and get, together with~\eqref{eq:PartDerivGeneralPattern}, the following:
\begin{equation}
  \sum_{n=0}^{\infty} P^{(k)}_{n} (x) t^{n} \Big|_{x = 1} = \frac{(2k-1)!! \hspace*{1pt} t^{k}}{\sqrt{1 - 2 t + t^{2}}^{2k+1}} = \frac{(2k-1)!! \hspace*{1pt} t^{k}}{\left| t - 1 \right|^{2k+1}} . \label{eq:GenFunctDerivRelationII}
\end{equation}
The left-hand-side of this equation reads
\begin{equation}
  \sum_{n=0}^{\infty} P^{(k)}_{n} (1) t^{n} \equiv \sum_{n=k}^{\infty} P^{(k)}_{n} (1) t^{n} = P^{(k)}_{k} (1) t^{k} + P^{(k)}_{k+1} (1) t^{k+1} + \ldots = t^{k} \times \left\{ P^{(k)}_{k} (1) + P^{(k)}_{k+1} (1) t + \ldots \right\} . \label{eq:LHSExpansion}
\end{equation}
Thus, we have to Taylor-expand the pole-term on the RHS of equation~\eqref{eq:GenFunctDerivRelationII}. We get (for expansion around $t_{0} = 0$):
\begin{align}
 \frac{1}{\left| t - 1 \right|^{2k+1}} &= \frac{1}{\left[ (t - 1)^{2} \right]^{(2k+1)/2}} = \left[ (t - 1)^{2} \right]^{-(2k+1)/2} \Big|_{t=0} + \frac{1}{1!} (-1) (2k+1) (t-1) \left[ (t - 1)^{2} \right]^{-(2k+3)/2} \Big|_{t=0} t  \nonumber \\
 & \hspace*{12.5pt} + \frac{1}{2!} (2k+1) (2k+2) \left[ (t - 1)^{2} \right]^{-(2k+3)/2} \Big|_{t=0} t^{2} \nonumber \\
 & \hspace*{12.5pt} + \frac{1}{3!} (-1) (2k+1) (2k+2) (2k+3) (t-1)  \left[ (t - 1)^{2} \right]^{-(2k+5)/2} \Big|_{t=0} t^{3} + \ldots  \nonumber \\
 &= 1 + \frac{1}{1!} (2k + 1) t + \frac{1}{2!} (2k + 1) (2k + 2) t^{2} + \frac{1}{3!} (2k + 1) (2k + 2) (2k + 3) t^{3} + \ldots \nonumber \\
 &= \sum_{n=0}^{\infty} \frac{(2k+n)!}{n! (2k)!} t^{n}  . \label{eq:TaylorExpPoleTerm}
\end{align}
Inserting this result into the relation~\eqref{eq:GenFunctDerivRelationII} and matching expressions, we get:
\begin{equation}
  P^{(k)}_{k+n} (1) = (2k - 1)!! \frac{(2k+n)!}{n! (2k)!} = (2k - 1)!! \frac{(k+[k+n])!}{([k+n]-k)! (2k)!} , \text{ for } n = 0,1,2,\ldots \label{eq:BoundaryValueResultI}
\end{equation}
or
\begin{equation}
  P^{(k)}_{n} (1) = (2k - 1)!! \frac{(n+k)!}{(2k)! (n-k)!} , \text{ for } n = k,k+1,k+2,\ldots .  \label{eq:BoundaryValueResultII}
\end{equation}
Some example-evaluations are:
\begin{align}
 P'_{n} (1) &= \frac{n (n+1)}{2}  , \label{eq:ExampleEvalResultI} \\
 P''_{n} (1) &= \frac{(n-1) n (n+1) (n+2)}{8}  , \label{eq:ExampleEvalResultI} \\
 P'''_{n} (1) &= \ldots  , \label{eq:ExampleEvalResultIII} 
\end{align}
which we observe to coincide with our test-evaluations previously done for the expression~\eqref{eq:BoundaryValuekthDerivativeAppII}. We can establish equality of the form of the result~\eqref{eq:BoundaryValueResultII} with our target-formula~\eqref{eq:BoundaryValuekthDerivativeAppII} by using the following known representation of the double-factorial:
\begin{equation}
 (2 k - 1)!! = \frac{(2k)!}{2^{k} k!}    . \label{eq:EquivRepDoubleFactorial}
\end{equation}
This relation can be quickly re-derived as follows: first, split $(2k)!$ into even and odd factors:
\begin{equation}
  (2k)! = (2k) \cdot (2k-1) \cdot (2k-2) \cdot \ldots \cdot 3 \cdot 2 \cdot 1  =  \left[ (2k) \cdot (2k - 2) \cdot \ldots \cdot 2  \right] \cdot \left[ (2k-1) \cdot (2k - 2) \cdot \ldots \cdot 1 \right]    .  \label{eq:DoubleFactReprProofI}
\end{equation}
Next, we can pull a factor of $2$ out of each factor occurring in the first bracket, i.e. for the even terms:
\begin{align}
  (2k)!  &=  \left[ (2 \cdot k) \cdot 2(k - 1) \cdot \ldots \cdot (2\cdot 1)  \right] \cdot \left[ (2k-1) \cdot (2k - 2) \cdot \ldots \cdot 1 \right] \nonumber \\
  &= 2^{k} \left[ k \cdot (k-1) \cdot \ldots \cdot 1 \right] \cdot \left[ (2k-1) \cdot (2k - 2) \cdot \ldots \cdot 1 \right] = 2^{k} k! (2k-1)!! .  \label{eq:DoubleFactReprProofII}
\end{align}
This concludes everything that had to be shown.


\end{document}